# Application of the projection operator formalism to non-Hamiltonian dynamics

Jianhua Xing[1, a], K. S. Kim[2]

1) Department of biological sciences, Virginia Polytechnic Institute and State University, Blacksburg, VA 24061
2) Lawrence Livermore National Laboratory and University of California, Livermore, CA
a) jxing@vt.edu

## Abstract:

Reconstruction of equations of motion from incomplete or noisy data and dimension reduction are two fundamental problems in the study of dynamical systems with many degrees of freedom. For the latter extensive efforts have been made but with limited success to generalize the Zwanzig-Mori projection formalism, originally developed for Hamiltonian systems close to thermodynamic equilibrium, to general non-Hamiltonian systems lacking detailed-balance. One difficulty introduced by such systems is the lack of an invariant measure, needed to define a statistical distribution. Based on a recent discovery that a non-Hamiltonian system defined by a set of stochastic differential equations can be mapped to a Hamiltonian system, we develop such general projection formalism. In the resulting generalized Langevin equations, a set of generalized fluctuation-dissipation relations connect the memory kernel and the random noise terms, analogous to Hamiltonian systems obeying detailed balance. Lacking of these relations restricts previous application of the generalized Langevin formalism. Result of this work may serve as the theoretical basis for further technical developments on model reconstruction with reduced degrees of freedom. We first use an analytically solvable example to illustrate the formalism and the fluctuation-dissipation relation. Our numerical test on a chemical network with end-product inhibition further demonstrates the validity of the formalism. We suggest that the formalism can find wide applications in scientific modeling. Specifically, we discuss potential applications to biological networks. In particular, the method provides a suitable framework for gaining insights into network properties such as robustness and parameter transferability.

# Introduction

A classical problem in dynamical systems theory is reconstructing equations of motion describing a complex system from incomplete or noisy data. One frequently encounters the following situation in studying many systems ranging from traffic flow to biological gene regulatory networks[1,2]: 1) there are time series data for a projection of the true states; 2) the process and measurement of the data are noisy; 3) the underlying dynamical law is not clear.

Closely related is finding methods for reducing the complexity of dynamical systems with many degrees of freedom, central to many problems in science and engineering. In general it is impractical, and often unnecessary, to track all the dynamical information of the whole system. A common practice involves the projection of the full dynamics of the whole system into that of its smaller subsystem resulting in some type of information contraction. The procedure leads to the celebrated Langevin and generalized Langevin dynamics. The Zwanzig-Mori (ZM) formalism is a formal procedure of projection, especially for Hamiltonian systems [3,4,5]. Some recent applications of the formalism include the work of Lange and Grubmüller on protein dynamics[6], that of Li on crystalline solids[7], and that of Shea and Oppenheim for a Brownian particle in a nonequilibrium bath [8]. Inspired by its great success in irreversible statistical mechanics, Chorin and coworkers, have suggested application of ZM formalism for higher-order optimal prediction methods for general dynamical systems[9]. As Zwanzig pointed out[5], while the formalism is generally applicable, its practical usefulness lies in defining an inner product (or measure) properly (see below for details). The choice is straightforward for a Hamiltonian system relaxing to equilibrium asymptotically, but not clear for a general non-thermal dynamical system, or for a Hamiltonian system driven out of equilibrium. Probability measure on the phase space is essential for the study of ensembles of solutions to general dynamical systems. The Liouville property of a volume preserving flow map in phase space (that found in Hamiltonian systems) is guaranteed by the Gibbs measure, an invariant measure of the flow map. For non-Hamiltonian systems (e.g. dissipative dynamical systems, nonequilibrium steady state systems, etc), the construction of an invariant phase space measure has been addressed by several researchers[10,11]. In their work [10], Tuckerman et al generalize the usual Hamiltonian based statistical mechanical phase space principles by examining how the phase space volume element transforms under arbitrary coordinate transformation at different times, thereby generalizing the Liouville flow through the introduction of a phase space metric. The resulting generalized equation of motion derived by this geometrical approach can serve as the starting point for further projection analysis. In this work, we focus on an alternative approach to non-Hamiltonian systems based on the stochastic dynamics point of view [12,13-15], which is complementary to the statistical mechanics viewpoint discussed above. In fact, the work of Graham [15] shares a number of interesting parallels with the geometrical approach of Tuckerman et al, but applied to Langevin and Fokker-Planck systems.

Recently one of us has proved that one can map a system described by a set of stochastic differential equations

$$\dot{x}_i = dx_i/dt = G_i(\mathbf{x}) + \sum_{j=1}^{M} g_{ij}(\mathbf{x})\zeta_j(t), \quad i=1,\cdots,N. \qquad \textbf{Eqn. 1}$$

to a Hamiltonian system [16]. In general the vector $\mathbf{G}(x)$ can not be represented as the gradient of a scalar potential due to violation of detailed balance, $M$ and $N$ may be different, $\zeta_i(t)$ are temporally uncorrelated, statistically independent Gaussian white noise with the averages satisfying $<\zeta_i(t)\zeta_j(\tau)> = \delta_{ij}\delta(t-\tau)$, $\mathbf{g(x)}$ is related to the $N \times N$ diffusion matrix $\mathbf{gg}^T = 2\mathbf{D}/\beta$, where the transpose of a matrix is designated by the superscript $T$. For a thermal system $\beta$ is the inverse temperature, $1/k_B T$ with $k_B$ the Boltzmann constant. For a non-thermal (non-Hamiltonian) system, $\beta$ is a parameter analogous to an effective inverse temperature. Eqn. 1 is widely used to describe dynamics in various fields of science from physics, ecology and cell biology, finance, geology, etc[12,17]. The mapping makes explicit the choice of the inner product definition, and thus derivation of the ZM projection formulae straightforward. The mapping is only used as an auxiliary tool to derive the formula. The actual construction of the mapping Hamiltonian is not needed.

The main purpose of this work is to formally derive the projection formula for a general dynamical system. While the derivation itself is rather involved, the final result can serve as the theoretical basis for optimal reconstruction of equations of motion in a reduced representation. In the remaining parts of the paper, we will first develop the theory, then present analytical results for a quasi-linear system and numerical tests on a small chemical network, and finally conclude with a general discussion.

# Theory

## *Summary of the Zwanzig-Mori formalism*

We follow the notation of Zwanzig here[5]. Consider a dynamical system whose time evolution is described by a Liouville equation in an *N* dimensional space,

$$\frac{\partial}{\partial t} R(\mathbf{x}(t),t) = LR(\mathbf{x}(t),t) \quad \textbf{Eqn. 2}$$

For a Hamiltonian system, the Liouville operator *L* is defined as,

$$LR = \sum_i \left( \frac{\partial H}{\partial p_i} \frac{\partial R}{\partial q_i} - \frac{\partial H}{\partial q_i} \frac{\partial R}{\partial p_i} \right)$$

where *H* is the Hamiltonian, and $q_i$ and $p_i$ are the coordinates and conjugate momenta. Any dynamical quantity $R(\mathbf{q}, \mathbf{p})$ is a vector in the Hilbert space. One can define the projection of another dynamical quantity *B* onto the subspace spanned by *R* as,

$$PB = \sum_{ij} (B, R_i)(R, R)^{-1}_{ij} R_j, \; i, j = 1, \cdots, l$$

where *l* is the number of components of the vector *R*. The inner product for two arbitrary variables *R* and *B* is defined as $(R, B) \equiv < R^\dagger B > \equiv \int R^\dagger B \psi(\mathbf{x}) d\mathbf{x}$, where † means conjugate transpose, and $\psi$ is a weighting function to be defined. Any dynamical variable within the subspace can be expressed as a linear combination of the basis functions. The projected equations of an arbitrary dynamical variable *R*, which is defined within the projected subspace, are in the form of generalized Langevin equations (GLEs),

$$\frac{\partial}{\partial t}R(\mathbf{x}(t),t) = PLR(\mathbf{x}(t),t) - \int_0^t ds\mathbf{K}(s)\cdot R(\mathbf{x}(t-s),t-s) + F(t) \qquad \text{Eqn. 3}$$

where $F(t) = \exp(t(\mathbf{1}-\mathbf{P})L)(\mathbf{1}-\mathbf{P})LR$ $\mathbf{K}(t) = -(LF(t),R)\cdot(R,R)^{-1}$. At time 0, $R$ is within the subspace. Time evolution of $R$ is decomposed into the dynamics within the subspace and within the orthogonal subspace, which are treated explicitly and implicitly respectively. Effects of the latter on the former are accounted for by the last two terms in the right hand side of Eqn. 3. If the Liouville operator is anti-Hermitian, one further obtains the generalized fluctuation-dissipation relation (GFDR) between the memory kernel and the random force term,

$$\mathbf{K}(t) = (F(t),LR)\cdot(R,R)^{-1} = (F(t),F(0))\cdot(R,R)^{-1}$$

Eqn. 3 is mathematically equivalent to Eqn. 2, and assumes a form formally analogous to the phenomenological generalized Langevin equation. Appendix A and B give some additional details of the projection formalism. In principle, one can apply the projection formalism to general dynamical systems, and the choice of the Hilbert subspace spanned by $R$ and the weighting functions can be arbitrary[5,18]. However, as Zwanzig pointed out[5], in general the procedure thus far is only a mathematical formalism with no obvious practical usage. For a concrete realization one has to choose $\psi$ properly. With an invariant measure, $\int f(\mathbf{x}(t))\psi(\mathbf{x}(0))d\mathbf{x}(0) = \int f(\mathbf{x}(t))\psi(\mathbf{x}(t))d\mathbf{x}(t) = \int f(\mathbf{x})\psi(\mathbf{x})d\mathbf{x}$, $R$ and $\mathbf{x}$ can be viewed as random variables[5,18]. For a Hamiltonian system near equilibrium, one can choose the Gibbs canonical distribution $\rho(\mathbf{q},\mathbf{p}) = \exp(-\beta H)/\int \exp(-\beta H)d\mathbf{q}d\mathbf{p}$ as $\psi$. The conservation of $H$ and the Liouville theorem ensures the choice is an invariant measure. The term **F**(t) has ensemble average <**F**(t)> = 0, and <**F**(t) R(0)> = 0, thus indeed behaves as a random force term. The memory kernel and the random force terms are not independent, but are constrained by the GFDR. For a general non-Hamiltonian system, one can either approach the problem by generalizing statistical mechanics via the formalism of Tuckerman et al [10], or by generalizing methods from stochastic systems which is then subsequently mapped to an equivalent Hamiltonian system [13,14,16]. In this work, we take the latter approach and leave the former to future studies. As Zwanzig has pointed out [5], the projection operator approach is most transparent in situations when there is a clear time scale separation between the slow primary and fast environmental degrees of freedom, although the GLE formalism has been applied to model systems with no clear time scale separation [19]. Phenomenological generalized Langevin equations have been used to model dynamical systems such as financial market fluctuations [20]. However, the lack of a GFDR makes model construction less defined[21], and imposes difficulties on the theoretical analysis of systems under colored noise[22].

## *Projection formalism for non-Hamiltonian systems*

In a recent paper[13], Ao shows that one can always construct a decomposition involving a symmetric matrix **S** and an anti-symmetric one **T** which transforms Eqn. 1 into,

$$\mathbf{M} \cdot \frac{d\mathbf{x}}{dt} = \mathbf{M} \cdot (\mathbf{G}(\mathbf{x}) + \mathbf{g}(\mathbf{x})\zeta(t)) = -\nabla_x \phi(\mathbf{x}) + \mathbf{g}'(\mathbf{x})\zeta(t) \qquad \text{Eqn. 4}$$

where, **M** = **S** + **T**, $\phi$ is a scalar function corresponding to the potential function in a Hamiltonian system satisfying $\mathbf{g}'\mathbf{g}'^T = 2\mathbf{S}/\beta$ and $(\partial \times \partial \phi)_i \equiv \sum_{j,k} \varepsilon_{ijk} \partial_j \partial_k \phi = 0$, where $\varepsilon_{ijk}$ is the Levi-Civita tensor. Then **S** and **T** are uniquely determined by

$\partial \times [\mathbf{M} \cdot \mathbf{G}(\mathbf{x})] = 0, (\mathbf{M})^{-1} + (\mathbf{M}^{-1})^T = 2\mathbf{gg}^T$, and proper choice of the boundary conditions.

The above equation can be augmented with an auxiliary momentum term (analogous to Andersen and Nosé extended system method[23]),

$$\dot{\mathbf{x}} = \frac{\tilde{\mathbf{p}}}{m} \qquad \text{Eqn. 5}$$

$$\dot{\mathbf{p}} = \left[ -\mathbf{T}(\mathbf{x}) \cdot \frac{\tilde{\mathbf{p}}}{m} - \nabla_x \phi(\mathbf{x}) \right] + \left[ -\mathbf{S}(x) \cdot \frac{\tilde{\mathbf{p}}}{m} + \mathbf{g}'(\mathbf{x}) \zeta(t) \right] \qquad \text{Eqn. 6}$$

which reduces to Eqn. 4 in the limit m→0, $\dot{\mathbf{p}} \to 0$.

One can verify the validity of Ao's decomposition procedure for (both linear and nonlinear) systems possessing steady-state (see Appendix C). It can be further shown that one can map the dynamics described by Eqn. 4 to a Hamiltonian system in the zero mass limit [16]. The proof proceeds in two steps. First define a Lagrangian so the resultant Euler-Lagrange equation gives Eqn. 5 and 6 excluding the dissipative terms (the terms inside the second bracket in Eqn. 6). Second following a procedure similar to that adopted by Zwanzig [24], one can replace the dissipative terms by a bath Hamiltonian with a large number of harmonic oscillators coupled to the primary degrees of freedom **x**. The bath is initially in contact with a heat reservoir, and the initial conditions of the bath degrees of freedom are drawn from a canonical distribution. The overall Hamiltonian is

$$H = \frac{(\mathbf{p} - \mathbf{A}(\mathbf{x}))^2}{2m} + \phi(\mathbf{x}) + \sum_{\alpha=1}^{N_\alpha} \left[ \sum_{j=1}^{N} \left( \frac{1}{2} p_{\alpha j}^2 + \frac{1}{2} \omega_{\alpha j}^2 (q_{\alpha j} - a_\alpha(\mathbf{x})/(\sqrt{N} \omega_{\alpha j}^2))^2 \right) \right] \qquad \text{Eqn. 7}$$

where **A** is a vector potential satisfying $T_{ij} = \partial A_i / \partial x_j - \partial A_j / \partial x_i$, $\mathbf{p} \equiv m\dot{\mathbf{x}} + \mathbf{A} = \tilde{\mathbf{p}} + \mathbf{A}$ is the conjugate momentum. The last term in Eqn. 7 is the bath Hamiltonian, and its form is determined by **S** [16]. The Hamiltonian describes a particle, coupled to a set of harmonic oscillators, moving in a hypothetical *n*-dimensional conservative scalar potential and magnetic (the vector potential) field. The mapping permits additional analytical methods from Hamiltonian dynamics to become available for dissipative non-Hamiltonian systems. For the current purpose, the property of Hamiltonian dynamics suggests the inner product definition,

$$(R,B) \equiv \frac{\int R^\dagger B \exp(-\beta H) d\mathbf{x} d\mathbf{p}}{\int \exp(-\beta H) d\mathbf{x} d\mathbf{p}}$$

Alternatively, one can also replace the integration over **p** by $\tilde{\mathbf{p}}$. Both definitions ensure the requirement of an invariant measure[25]. As will be clear from the following theoretical developments and examples, in real applications one need not perform the mapping. The mapping merely serves as an auxiliary tool to derive the projection formulae and the GFDR.

For simplicity we only consider projecting to a subspace composed of the first *l* components of **x** and the corresponding velocity components. Generalization to collective coordinates is straightforward, and is given in Appendix D (see also refs [4,6]). In the following discussion we denote them as $\mathbf{X} = \{x_1, x_2, \ldots, x_l\}$, and $\dot{\mathbf{X}} = \{\dot{x}_1, \dot{x}_2, \ldots, \dot{x}_l\}$. Let's define a nonlinear projection operator,

$$Ph = \frac{1}{\bar{\rho}(\bar{\mathbf{X}}, \dot{\bar{\mathbf{X}}})} \int h\rho(\mathbf{x},\mathbf{p})\delta(\bar{\mathbf{X}}-\mathbf{X})\delta(\dot{\bar{\mathbf{X}}}-(\mathbf{p}-\mathbf{A})/m)d\mathbf{x}d\mathbf{p}$$

where $h$ is an arbitrary function, and $\bar{\rho}(\bar{\mathbf{X}}, \dot{\bar{\mathbf{X}}}) = \int \rho(\mathbf{x},\mathbf{p})\delta(\bar{\mathbf{X}}-\mathbf{X})\delta(\dot{\bar{\mathbf{X}}}-(\mathbf{p}-\mathbf{A})/m)d\mathbf{x}d\mathbf{p}$.

Then (see Appendix D for detailed derivation),

$$PLX_j = \dot{X}_j \qquad \text{Eqn. 8}$$

$$PL\dot{X}_j = -\frac{1}{m\beta}\frac{\partial}{\partial X_j}\ln\bar{\rho}(\mathbf{X}) - \sum_{k\neq j}\dot{X}_k\left(<\frac{\partial A_k}{\partial X_j}>_{\mathbf{x}} - <\frac{\partial A_j}{\partial X_k}>_{\mathbf{x}}\right) \qquad \text{Eqn. 9}$$

where,

$$\bar{\rho}(\bar{\mathbf{X}}) = \int d\mathbf{x}\exp(-\beta\phi)\delta(\bar{\mathbf{X}}-\mathbf{X})$$

$$<B>_{\bar{\mathbf{x}}} = \frac{1}{\bar{\rho}(\bar{\mathbf{X}})}\int d\mathbf{x}B(\mathbf{x})\exp(-\beta\phi)\delta(\bar{\mathbf{X}}-\mathbf{X})$$

and we have omitted the bar on the variables. The projected equation of motion is,

$$m\frac{\partial^2}{\partial t^2}X_j(t) = -\frac{\partial}{\partial X_j}W(\mathbf{X}) + \sum_i\left[\bar{T}_{ij} + \int_0^t dsmK_{ji}(s)\dot{X}_i(t-s)\right] + F_j(t) \qquad \text{Eqn. 10}$$

where $W(\mathbf{X}) = -\ln\bar{\rho}(\mathbf{X})/\beta$ is the potential of mean force, and $\bar{T}_{ij} = <\partial A_i/\partial x_j - \partial A_j/\partial x_i>_{\mathbf{x}}$ is the renormalized antisymmetric matrix in the reduced space. A generalized fluctuation-dissipation relation relates the memory kernel and the random force,

$$<F_i(t)F_j(t')> = mK_{ij}(t-t')/\beta \qquad \text{Eqn. 11}$$

Taking the ansatz $mK_{ji}(t) = (2\gamma_{ji}^0 \delta(t) + \gamma_{ji}(t))$ [6], and the zero-mass limit, one has,

$$0 = \left[ -\frac{\partial}{\partial X_j} W(\mathbf{X}) - \sum_i (\bar{S}_{ji} + \bar{T}_{ji}) \dot{X}_i(t) \right] - \left[ \sum_i \int_0^t ds \gamma_j(t-s) \dot{X}(s) \right] + F_j(t) \quad \text{Eqn. 12}$$

where $\bar{S}_{ij} = \gamma_{ij}^0$, and the matrices $\bar{\mathbf{S}}$ and $\bar{\mathbf{T}}$ are $\mathbf{X}$-dependent in general. Eqn. 12 is in the form of the generalized Langevin equation (GLE), and together with the GFDR Eqn. 11 is the main result of this work. The form of the equation is similar to what is used for thermal systems, except for the presence of an extra anti-symmetric matrix $\bar{\mathbf{T}}$ reflecting that the system violates detailed balance. This result can be viewed as a generalization of the so-called "geometric magnetism" [26]. Physically the terms in the first brackets are related to the direct interactions (or fluxes) within the projected subspace and between the subspace and the surroundings. The terms in the second brackets refer to the retarded interactions mediated by the implicit surroundings. The two terms are analogous to the adiabatic Born-Oppenheimer potential and the nonadiabatic corrections in molecular physics. Note that the complicated nonlinear transformation prescribed by Ao and by Xing[13,16] is not needed.

For the case where we project to a 1-D system, the equation is,

$$0 = -\frac{\partial}{\partial X_j} W(\mathbf{X}) - \int_0^t ds \Gamma(s) \dot{X}(t-s) + m\beta F(t) \quad \text{Eqn. 13}$$

with $\Gamma(t) = 2\gamma_0 \delta(t) + \gamma_1(t)$

# Analytical and numerical examples

## *Linear example*

Here we use a two-dimensional system to demonstrate how the transformation reveals some hidden relation between the memory kernel and the noise term,

$$\dot{x}_1 = -a_{11}x_1 - a_{12}x_2 + g_1\zeta_1(t)$$
$$\dot{x}_2 = -a_{21}x_1 - a_{22}x_2 + g_2\zeta_2(t)$$

Eqn. 14

For this and the following numerical examples, we choose units where $\beta = 1$ for simplicity. First, let's integrate out subsystem 2 directly. Notice that,

$$x_2(t) = \exp(-a_{22}t)x_2(0) + \int_0^t \exp(-a_{22}\tau)(-a_{21}x_1(t-\tau) + g_2\zeta_2(t-\tau))d\tau$$
$$= \exp(-a_{22}t)\left(x_2(0) + \frac{a_{21}}{a_{22}}x_1(0)\right) - \frac{a_{21}}{a_{22}}x_1(t) + \int_0^t \exp(-a_{22}\tau)(-\frac{a_{21}}{a_{22}}\dot{x}_1(t-\tau) + g_2\zeta_2(t-\tau))d\tau$$

Then

$$\dot{x}_1 = -a_{11}x_1 + \int_0^t \{\exp(-a_{22}\tau)a_{12}a_{21}\}x_1(t-\tau)d\tau$$
$$+ \left\{-a_{12}\exp(-a_{22}t)x_2(0) + g_1\zeta_1(t) - \int_0^t \exp(-a_{22}\tau)a_{12}g_2\zeta_2(t-\tau))d\tau\right\}$$
$$= -\left(a_{11} - \frac{a_{12}a_{21}}{a_{22}}\right)x_1(t) + \int_0^t \left\{\exp(-a_{22}\tau)\frac{a_{12}a_{21}}{a_{22}}\right\}\dot{x}_1(t-\tau)d\tau$$
$$+ \left\{-a_{12}\exp(-a_{22}t)\left(x_2(0) + \frac{a_{21}}{a_{22}}x_1(0) - \int_0^t \exp(-a_{22}\tau)a_{12}g_2\zeta_2(t-\tau)d\tau\right) + g_1\zeta_1(t)\right\}$$

Eqn. 15

If one identifies the first term as the derivative of the potential of mean force, and the two terms in the curly brackets as the memory and noise terms, the above equation formally resembles the generalized Langevin equation. However, the relation between the memory kernel and the noise is not transparent in any of these two expression in Equation 15.

Let us transform Eqn 14 by multiplying $\mathbf{M} = \mathbf{S} + \mathbf{T}$ on both sides,

$$0 = -M_{11}\dot{x}_1 - M_{12}\dot{x}_2 - \phi_{11}x_1 - \phi_{12}x_2 + \xi_1(t)$$
$$0 = -M_{21}\dot{x}_1 - M_{22}\dot{x}_2 - \phi_{12}x_1 - \phi_{22}x_2 + \xi_2(t)$$

**Eqn. 16**

where, $\phi = \mathbf{M} \cdot \mathbf{a}, \xi = \mathbf{M} \cdot \mathbf{g}\zeta$. The matrix $\phi$ is symmetric, and one can obtain the analytical expression of $\mathbf{M}$ following the procedure by Kwon et al.[14]

Therefore,

$$0 = \left\{\left(-\phi_{11} + \frac{\phi_{12}\phi_{12}}{\phi_{22}}\right)x_1(t)\right\} - \int_0^t d\tau \left\{\frac{2}{m}\left(M_{11} - \frac{M_{12}M_{21}}{M_{22}}\right)\delta(t)\right.$$
$$+ \frac{\phi_{22}}{m}e^{-\frac{\phi_{22}}{M_{22}}\tau}\left(\frac{M_{12}}{M_{22}} - \frac{\phi_{12}}{\phi_{22}}\right)\left(\frac{M_{21}}{M_{22}} - \frac{\phi_{21}}{\phi_{22}}\right)\right\}p_1(t-\tau) + \left\{e^{-\frac{\phi_{22}}{M_{22}}t}\left(\frac{M_{12}}{M_{22}} - \frac{\phi_{12}}{\phi_{22}}\right)(\phi_{21}x_1(0) + \phi_{22}x_2(0))\right.$$
$$+ \xi_1(t) - \frac{M_{12}}{M_{22}}\xi_2(t) + \int_0^t d\tau e^{-\frac{\phi_{22}}{M_{22}}\tau}\frac{\phi_{22}}{M_{22}}\left(\frac{M_{12}}{M_{22}} - \frac{\phi_{12}}{\phi_{22}}\right)\xi_2(t-\tau)\right\}$$

**Eqn. 17**

Comparing the above final expression with Eqn. 13, one can identify the terms in the three curly brackets corresponding to the derivative of the potential of mean force, the memory term, and the random force term.

Note that while Eqns. 14 and 16 differ only by a matrix transformation, there is no simple relation between Eqns. 15 and 17.

In Appendix E, we demonstrated the validity of the GFDR for this system. In Appendix F, we derive analytical expressions of the GLEs for a general (N+1)-quasi-linear (in the transformed representation) dimensional system. This model can be regarded as a generalization of the well studied system-bath model in Hamiltonian systems[5].

## *Numerical example*

In most applications of the projection formalism including previous work on Hamiltonian systems, it is impractical to perform the projection analytically. Extensive studies of related methods exist for Hamiltonian systems, which can be generalized to non-Hamiltonian systems by the above mapping. We will leave these for future studies. Here for illustrative purposes, we will demonstrate the validity and utility of the projection method by applying it to a simple chemical network. The network is an end-product inhibition motif commonly found in metabolic and other biological regulatory networks (see Fig. 1a)[27]. Each reaction is governed by irreversible Michaelis-Menten kinetics,

$$\dot{x}_1 = \frac{v_m}{K_m + x_4} - \frac{v_m x_1}{K_m + x_1} + g\zeta_1(t)$$

and similar expressions for other species concentrations. The values $v_m = 1$, $K_m = 0.5$, $g = 0.005$ are used in the simulations. Numerical details are given in Appendix G.

The system is initially at the steady-state. At time 0, the concentration of $x_1$ is set to a value $x_1(0)$. The relaxation dynamics $\bar{x}_1(t)$, which is defined as the value of $x_1$ averaged over all trajectories, is monitored. We use the result of one simulation on the full model, along with $x_1(0) = 2 < x_1 >_{ss}$, as the known information to fit the parameters for the GLE

Eqn. 13. The term $<x_1>_{ss}$ refers to the steady-state average. Fig 1b-d show the fitting results and parameters. The potential of mean force is calculated from the steady-state distribution. We model the nonsingular part of the memory kernel, $\gamma_1$ with a Gaussian basis set. Physically one may understand the memory kernel fitting as follows: the change of $x_1$ is propagated to $x_4$ through a series of reactions, and acts back on $x_1$ at a later time. The first portion of $\gamma_1$ with negative values (as modeled by two Gaussian functions) accounts for most of the effect. Some remnant effect propagates one more cycle to act on $x_1$ with an opposite sign (inhibition of inhibition), doubled delay time, and reduced amplitude. The focus of this work is to illustrate the validity of the method, and thus makes no effort to fully optimize the fitting. We then use the set of parameters to simulate the GLE with different values of $x_1(0)$, and compared with simulation results of the full model. Fig 2 shows remarkable agreement in three cases. Even without fully optimizing the parameters, the GLE model can reproduce the damped oscillating temporal behaviors of $\bar{x}_1(t)$. These results support the validity of Eqn. 13.

## Discussions and concluding remarks

In this work we develop a generalized Zwanzig-Mori projection formalism for dissipative non-Hamiltonian systems lacking detailed-balance. Because of the mapping between a dissipative non-Hamiltonian system and a Hamiltonian system, we expect that the large number of existing methods applying the projection method to Hamiltonian systems can be readily applied to non-Hamiltonian systems[21,28]. We suggest that an important direction for future research is to develop a number of standard ansatz (functional forms for the potential of mean force, the memory kernel, etc) for different situations. Analytical results for the quasi-linear systems may serve as starting points.

In this work we mainly focus on theoretical method development, and leave further technical developments and applications for future studies. The method is suitable for a system with stationary distribution, while some standard techniques exist to render certain nonstationary processes stationary [2]. Assuming that the primary degrees of freedom have been identified in a real application, we suggest the following generic procedure: first, one constructs the potential of mean force from the stationary distributions; second, one reconstructs the memory kernel and the anti-symmetric matrix from the time series data [2,21]; the noise term is automatically given by the GFDR. One should notice that in previous usages of the GLE formalism for systems without detailed balance [20,21], due to the absence of a GFDR, the term corresponding to the potential of mean force in this work is not directly related to the stationary distribution, and the memory and noise terms have to be determined separately. Several studies have discussed the application of the Zwanzig-Mori projection to time series data analysis [29]. However, our results are not limited to linear analysis and systems satisfying detailed balance.

In most applications of the projection formalism including previous work on Hamiltonian systems, it is often impractical to perform the projection procedure analytically. For heuristic purposes, we test the formalism by projecting the systems with two and three dimensions to one dimension. The formalism, however, can be applied to arbitrary high dimensional systems. In Appendix E, we derived the analytical formula for projecting an ($N$+1)-dimensional quasi-linear system to one dimension. In real applications, information about a system is often incomplete. Instead one can obtain the potential of mean force, the matrices $\bar{\mathbf{S}} + \bar{\mathbf{T}}$ and $\gamma$ from available data, following the well-established procedures developed for Hamiltonian systems[6,21,28]. On multiplying $(\bar{\mathbf{S}} + \bar{\mathbf{T}})^{-1}$ to both sides of Eqn. 12, one transforms back to the original representation having direct physical meaning for each term.

Given the broad range of problems Eqn. 1 describes, we expect that the method discussed in this work may yield fruitful applications for a number of scientific disciplines. Here we will discuss its usage and implications in the field of mathematical modeling of biological networks, or systems biology in a broader sense.

On modeling a complex dynamical system, a common problem is that there is insufficient information to identify a large number of parameter values in the model. For example in the field of systems biology, one frequent criticism of mathematical modeling lies in methods attempting to fit several data points with dozens or even hundreds of parameters. Fortunately analysis shows that for many systems the quality of the data fitting is usually largely affected by a small number of composite parameters, and insensitive to others[30]. A mathematical model with many variables and parameters is also computationally expensive. The present projection formalism provides a systematic method to construct a reduced model with a small number of variables and parameters important for the dynamics under study. It also provides a method for performing multi-scale modeling, using information obtained from finer level simulations for constructing a coarse-grained model.

Network robustness is a related problem. It has been suggested that robustness is a general property for many biological networks. As illustrated by Barkai and Leibler using the bacterial chemotaxis network model, a system is robust if its function is determined by one or a small number of composite quantities, and values of the latter are insensitive to variation of most control parameter[31]. The projection method provides a natural framework for quantifying network robustness under perturbations

There are extensive discussions on whether one can use the parameters measured in vitro on modeling processes *in vivo*[32]. For the latter, there are inevitably interactions between the subsystem one examines and the remaining part of the living system, which are not present in the system *in vitro*. The projection method provides a theoretical explanation why in general the two sets of parameters should be different. Even in the case when the memory kernel can be approximated by a delta function (so the retarded memory term $\gamma_{ij}$ = 0), the interactions between the subsystem that one models and the remaining degrees of freedom affect the dynamics of the subspace through the renormalization of the model parameters[33]. The projection method can suggest a controlled approximation linking the two sets of parameters.

We thank Drs. Oliver Lange, Helmut Grubmüller, Attila Szabo, Katja Lindenberg, and Michael Surh for discussions, and Ms Yan Fu for making Fig 1a. JX is supported by NSF (grant number DMS-0969417).

# Appendix

## *A. Linear and nonlinear projection*

As Zwanzig points out, the term "nonlinear projection" is actually misleading. Both linear and nonlinear projections work within a framework of a linear Hilbert space. They differ only by the size of the Hilbert subspace. The formal theoretical development discussed in the section "Summary of the Zwanzig-Mori formalism" applies for both cases. As an example, let us consider a 1-D Hamiltonian system. If one chooses the subspace as expanded by $\{x_1, p_1\}$ (the quantity $R$ in the main text), the projection is linear. The projection is called "nonlinear" if the basis is expanded by including higher order functions of $x_1$. Zwanzig presents a choice of the subspace by including all the possible functional forms of the projected degrees of freedom. The Hilbert subspace has infinite dimension. For this specific example, the basis is composed of $\{x_1, x_1^2, ...; p_1\}$.

## *B. About the term* $(R,R)^{-1}$

Both the memory kernel and the GFDR contain the term $(R,R)^{-1}$. Here we will use the above mentioned 1-D system to derive the needed elements of $(R,R)^{-1}$. Generalization to multi-dimensions is straightforward. For a Hilbert space with basis $\{x_1, x_1^2, ..., x_1^n, p_1\}$, notice that $(x_1^i, p_1) = 0$. Therefore, the matrix (R, R) has the block form,

$$(R,R) = \begin{pmatrix} (x_1,x_1) & \cdots & (x_1^n,x_1^n) & 0 \\ & \cdots & & \\ (x_n,x_1) & \cdots & (x_n,x_n) & 0 \\ 0 & \cdots & 0 & (p_1,p_1) \end{pmatrix}$$

The momentum-containing term is block diagonal. In the projected equations, one needs only the momentum-containing term $((R,R)^{-1})_{n+1,n+1}$, which is $(p_1,p_1)^{-1} = \beta/m$.

## C. Mapping between non-Hamiltonian and Hamiltonian systems

We use **p** and $\tilde{\mathbf{p}}$ to represent the conjugate momentum and the kinetic momentum, respectively. That is, we switch the notations **p** and $\tilde{\mathbf{p}}$ as in ref.[16] to simplify our discussion since we use the conjugate momentum in most derivations.

The decomposition introduced by Ao is not transparent, especially to nonlinear systems. By working with the corresponding Fokker-Planck equations, Xing, shows that one can relate the procedure of Ao to other well-established result, and thus prove its validity [16]. Below we summarize the result, with details found in the recently published paper [16].

For the expanded equations 5 and 6, Yin and Ao[34] show that the corresponding Fokker-Planck equation is $\partial_t \rho = \nabla \cdot \mathbf{M}^{-1}[\nabla + \nabla \phi]\rho$, which can be further rewritten as,

$\partial_t \rho = \nabla \cdot [\mathbf{D} \cdot \nabla + (\nabla \cdot \mathbf{Q}^T) + (\mathbf{D} + \mathbf{Q}) \cdot \nabla \phi]\rho$, where $\mathbf{M}^{-1} = (\mathbf{S} + \mathbf{T})^{-1} = \mathbf{D} + \mathbf{Q}$, with the symmetric matrix $\mathbf{D} = 2\beta \mathbf{g}\mathbf{g}^T$, and **Q** an anti-symmetric matrix. To derive the above expression we have used the antisymmetric property of **Q**, and noticed,

$$\nabla \cdot (\mathbf{Q} \cdot \nabla \rho) = (\nabla \cdot \mathbf{Q}) \cdot (\nabla \rho) = \sum_{ij} \partial_j \left( \frac{\partial Q_{ij}}{\partial x_i} \rho \right) = -\sum_{ij} \partial_j \left( \frac{\partial Q_{ji}}{\partial x_i} \rho \right) = \nabla \cdot (\nabla \cdot \mathbf{Q}^T) \rho \qquad \text{Eqn. 18}$$

Alternatively, let us closely follow a procedure used by Graham[15] and by Eyink et al.[35] on a generic Fokker-Planck equation,

$$\partial_t \rho(\mathbf{x},t) = \nabla \cdot [-\mathcal{J} \rho(\mathbf{x},t) + \mathbf{D} \cdot \nabla \rho(\mathbf{x},t)] \qquad \text{Eqn. 19.}$$

The explicit form of $\mathcal{J}$ depends on the choice of interpretations of the SDEs. One can define an "entropy" term from the steady state distribution $\mathcal{S} \equiv \ln(\rho_{ss})$. Notice that one can decompose the flux term $\mathcal{J}$ in Eqn. 19 into a conservative part $\mathcal{R}$ and a dissipative part $\mathcal{D}$, $\mathcal{D} \equiv \mathbf{D} \cdot \nabla \mathcal{S} = -\mathbf{L}^s \cdot \mathbf{X}, \mathcal{R} \equiv \mathcal{J} - \mathcal{D}$. In the above expression we have written in the Onsager form with $\mathbf{L}^s \equiv \mathbf{D}$ being a symmetric matrix, and $\mathbf{X} \equiv -\nabla \mathcal{S}$ being the general force. Substituting the expression of $\rho_{ss}$ into Eqn. 19, one has, $\nabla \cdot [\mathcal{R} \exp(\mathcal{S})] = 0$. Then following Graham, and Eyink *et al.*, one can relate the divergence free vector $\mathcal{R} \exp(\mathcal{S})$ to an antisymmetric matrix $\mathcal{F}$, $\mathcal{R}_i \exp(\mathcal{S}) \equiv \sum_j \frac{\partial \mathcal{F}_{ij}}{\partial x_j}$. The antisymmetric matrix $\mathcal{F}$ can be further expressed in terms of a vector potential (see below). Therefore,

$$\mathcal{R}_i = L^a_{ij} \cdot \frac{\partial \mathcal{S}}{\partial x_j} + \frac{\partial L^a_{ij}}{\partial x_j}$$

with the new antisymmetric matrix $L^a_{ij} \equiv \mathcal{F}_{ij} \exp(-\mathcal{S})$. Therefore the drift term $\mathcal{J}$ can be expressed as,

$$\mathcal{J}_i = -(L^s_{ij} + L^a_{ij}) X_j + \sum_j \frac{\partial L^a_{ij}}{\partial x_j} \qquad \text{Eqn. 20}$$

We essentially reproduced the derivation of Eyink *et al.* but without the need for the divergence free approximation on $\mathbf{L}^s$ and $\mathbf{L}^a$. Graham discusses how one can construct the matrix $\mathcal{F}$ and so $\mathbf{L}^a$ from the steady state distribution and current with additional constraints (e.g., satisfying homegeneous Maxwell' equation)[15].

Up to now we have not specified the choice of $\mathcal{J}$. For a special choice $\mathcal{J}_{Ao} = \mathbf{G} - \nabla \cdot (\mathbf{L}^a)^T$, Eqn. 20 gives $\mathbf{G} = -(\mathbf{L}^s + \mathbf{L}^a) \cdot \mathbf{X}$. The subscript of $\mathcal{J}$ indicates that this choice corresponds to the zero-mass limit interpretation of Ao[13], with $\mathcal{S} \equiv -\phi$, $\mathbf{L}^s \equiv \mathbf{D}$, and $\mathbf{L}^a \equiv \mathbf{Q}$. In the above derivations, there is no reference to linearity of the system. Therefore the decomposition is valid for nonlinear systems.

Next one can define a Langrangian,

$$\mathcal{L}_0 = \frac{1}{2}m\dot{\mathbf{x}}^2 - \phi + \dot{\mathbf{x}} \cdot \mathbf{A}(\mathbf{x})$$

The Euler-Lagrange equation of $\delta \mathcal{L}_0 = 0$,

$$\frac{d}{dt}\left(\frac{\partial \mathcal{L}_0}{\partial \dot{x}_i}\right) - \frac{\partial \mathcal{L}_0}{\partial x_i} = 0$$

results in Eqns. 5 and 6 without the dissipative terms in Eqn. 6, provided,

$$T_{ij} = \frac{\partial A_i}{\partial x_j} - \frac{\partial A_j}{\partial x_i}$$

Therefore, $\mathbf{T}$ is analogous to the electromagnetic tensor[36]. One can also define the conjugate momentum,

$$\tilde{p}_i = \frac{\partial \mathcal{L}_0}{\partial \dot{x}_i} = p_i + A_i(\mathbf{x})$$

and a corresponding Hamiltonian,

$$\mathcal{H}_0 = \tilde{\mathbf{p}} \cdot \dot{\mathbf{x}} - \mathcal{L}_0 = \frac{(\tilde{\mathbf{p}} - \mathbf{A}(\mathbf{x}))^2}{2m} + \phi(\mathbf{x})$$

Next following Zwanzig[24], one can show that the dissipative terms in Eqn. 6 can be replaced by a harmonic bath Hamiltonian. Consequently, a general dissipative system described by stochastic Langevin equations and possessing steady states can be mapped to a Hamiltonian system.

## D. Derivation of the projection formula

Here we derive a more general projection formula, which reduces to Eqns 8 and 9. The procedure resembles that of Lange and Grubmüller [6].

Let's suppose that we project to a manifold,

$$\mathbf{c} = \mathbf{f}(\mathbf{x}), \dot{\mathbf{c}} = \nabla \mathbf{f}(\mathbf{x}) \cdot \dot{\mathbf{x}} = \nabla \mathbf{f}(\mathbf{x}) \cdot (\mathbf{p} - \mathbf{A}) \frac{1}{m}$$

The Liouville operator is

$$L = \sum_{i=1}^{N} \left( \frac{\partial H}{\partial p_i} \frac{\partial}{\partial x_i} - \frac{\partial H}{\partial x_i} \frac{\partial}{\partial p_i} \right) + \sum_{i=1}^{N} \sum_{\alpha=1}^{N_\alpha} \left( \frac{\partial H}{\partial p_{i\alpha}} \frac{\partial}{\partial q_{i\alpha}} - \frac{\partial H}{\partial q_{i\alpha}} \frac{\partial}{\partial p_{i\alpha}} \right)$$

Notice that $c_j$ and $\dot{c}_j$ have no explicit dependence on the bath variables. We define the projection operator as,

$$Ph = \frac{1}{\bar{\rho}(\mathbf{c},\dot{\mathbf{c}})} \int h\rho(\mathbf{x},\mathbf{p})\delta(\mathbf{c}-\mathbf{f})\delta(\dot{\mathbf{c}}-\nabla\mathbf{f}(\mathbf{x})\cdot\dot{\mathbf{x}})d\mathbf{x}d\mathbf{p}$$

where $h$ is an arbitrary function, and

$$\bar{\rho}(\mathbf{c},\dot{\mathbf{c}}) = \int \rho(\mathbf{x},\mathbf{p})\delta(\mathbf{c}-\mathbf{f})\delta(\dot{\mathbf{c}}-\nabla\mathbf{f}(\mathbf{x})\cdot\dot{\mathbf{x}})d\mathbf{x}d\mathbf{p}$$

The projection of $c_j$ is simple since it is still within the subspace,

$$Lc_j = \sum_{i=1}^{N}\left(\frac{\partial H}{\partial p_i}\frac{\partial c_j}{\partial x_i} - \frac{\partial H}{\partial x_i}\frac{\partial c_j}{\partial p_i}\right) = \sum_i \frac{\partial H}{\partial p_i}\frac{\partial f_j}{\partial x_i} = \dot{c}_j$$

The projection of $\dot{c}_j$ is given by,

$$PL\dot{c}_j = \sum_{i=1}^{N}\frac{1}{\bar{\rho}(\mathbf{c},\dot{\mathbf{c}})}\int d\mathbf{x}d\mathbf{p}\delta(\mathbf{c}-\mathbf{f})\delta(\dot{\mathbf{c}}-\nabla\mathbf{f}(\mathbf{x})\cdot\dot{\mathbf{x}})\exp(-\beta H)\left(\frac{\partial H}{\partial p_i}\frac{\partial \dot{c}_j}{\partial x_i} - \frac{\partial H}{\partial x_i}\frac{\partial \dot{c}_j}{\partial p_i}\right)$$

$$= \sum_{i=1}^{N}-\frac{1}{\beta\bar{\rho}(\mathbf{c},\dot{\mathbf{c}})}\int d\mathbf{x}d\mathbf{p}\delta(\mathbf{c}-\mathbf{f})\delta(\dot{\mathbf{c}}-\nabla\mathbf{f}(\mathbf{x})\cdot\dot{\mathbf{x}})\left(\frac{\partial}{\partial p_i}\exp(-\beta H)\frac{\partial \dot{c}_j}{\partial x_i} - \frac{\partial}{\partial x_i}\exp(-\beta H)\frac{\partial \dot{c}_j}{\partial p_i}\right)$$

$$= \sum_{i=1}^{N}\frac{1}{\beta\bar{\rho}(\mathbf{c},\dot{\mathbf{c}})}\int \left(\exp(-\beta H)\frac{\partial c_j}{\partial x_i}\right)\frac{\partial}{\partial p_i}\left[\delta(\mathbf{c}-\mathbf{f})\delta(\dot{\mathbf{c}}-\nabla\mathbf{f}(\mathbf{x})\cdot(\mathbf{p}-\mathbf{A})\frac{1}{m})\right]d\mathbf{x}d\mathbf{p}$$

$$-\sum_{i=1}^{N}\frac{1}{\beta\bar{\rho}(\mathbf{c},\dot{\mathbf{c}})}\int \left(\exp(-\beta H)\frac{\partial c_j}{\partial p_i}\right)\frac{\partial}{\partial x_i}\left[\delta(\mathbf{c}-\mathbf{f})\delta(\dot{\mathbf{c}}-\nabla\mathbf{f}(\mathbf{x})\cdot(\mathbf{p}-\mathbf{A})\frac{1}{m})\right]d\mathbf{x}d\mathbf{p}$$

$$= \sum_{i=1}^{N} \frac{1}{\beta \bar{\rho}(\mathbf{c},\dot{\mathbf{c}})} \int d\mathbf{x} d\mathbf{p} \exp(-\beta H) \left( \frac{1}{m} \frac{\partial}{\partial x_i} (\nabla f_j(\mathbf{x})) \cdot (\mathbf{p}-\mathbf{A}) - \frac{1}{m} \nabla f_j(\mathbf{x}) \cdot \frac{\partial}{\partial x_i} \mathbf{A} \right)$$

$$\left\{ \sum_k \delta(\mathbf{c}-\mathbf{f}) \frac{1}{m} \frac{\partial f_k}{\partial x_i} \frac{\partial}{\partial \dot{c}_k} \left[ \delta(\dot{\mathbf{c}} - \nabla \mathbf{f}(\mathbf{x}) \cdot (\mathbf{p}-\mathbf{A}) \frac{1}{m}) \right] \right\} - \sum_{i=1}^{N} \frac{1}{\beta \bar{\rho}(\mathbf{c},\dot{\mathbf{c}})} \int d\mathbf{x} d\mathbf{p} \exp(-\beta H) \frac{1}{m} \frac{\partial f_j}{\partial x_i}$$

$$\left\{ \sum_k \frac{\partial f_k}{\partial x_i} \frac{\partial}{\partial c_k} \left[ \delta(\mathbf{c}-\mathbf{f}) \delta(\dot{\mathbf{c}} - \nabla \mathbf{f}(\mathbf{x}) \cdot (\mathbf{p}-\mathbf{A}) \frac{1}{m}) \right] \right\} - \sum_{i=1}^{N} \frac{1}{\beta \bar{\rho}(\mathbf{c},\dot{\mathbf{c}})} \int d\mathbf{x} d\mathbf{p} \exp(-\beta H) \frac{1}{m} \frac{\partial f_j}{\partial x_i}$$

$$\left\{ \sum_k \frac{\partial}{\partial x_i} \left( \frac{1}{m} \nabla f_k \cdot (\mathbf{p}-\mathbf{A}) \right) \frac{\partial}{\partial \dot{c}_k} \left[ \delta(\mathbf{c}-\mathbf{f}) \delta(\dot{\mathbf{c}} - \nabla \mathbf{f}(\mathbf{x}) \cdot (\mathbf{p}-\mathbf{A}) \frac{1}{m}) \right] \right\}$$

$$= -\frac{1}{\beta \bar{\rho}(\mathbf{c},\dot{\mathbf{c}})} \sum_k \frac{\partial}{\partial c_k} \int d\mathbf{x} d\mathbf{p} \exp(-\beta H) \left[ \delta(\mathbf{c}-\mathbf{f}) \delta(\dot{\mathbf{c}} - \nabla \mathbf{f}(\mathbf{x}) \cdot (\mathbf{p}-\mathbf{A}) \frac{1}{m}) \right] \left\{ \sum_i \frac{1}{m} \frac{\partial f_j}{\partial x_i} \frac{\partial f_k}{\partial x_i} \right\}$$

$$-\frac{1}{\beta \bar{\rho}(\mathbf{c},\dot{\mathbf{c}})} \sum_{k \neq j} \frac{\partial}{\partial \dot{c}_k} \int d\mathbf{x} d\mathbf{p} \exp(-\beta H) \left[ \delta(\mathbf{c}-\mathbf{f}) \delta(\dot{\mathbf{c}} - \nabla \mathbf{f}(\mathbf{x}) \cdot (\mathbf{p}-\mathbf{A}) \frac{1}{m}) \right]$$

$$\frac{1}{m} \sum_i \left\{ \frac{\partial f_j}{\partial x_i} \frac{\partial}{\partial x_i} \left( \frac{1}{m} \nabla f_k \cdot (\mathbf{p}-\mathbf{A}) \right) - \frac{\partial f_k}{\partial x_i} \frac{\partial}{\partial x_i} \left( \frac{1}{m} \nabla f_j \cdot (\mathbf{p}-\mathbf{A}) \right) \right\}$$

**Eqn. 21**

To derive the above expression, we performed integration by parts, and used the relations,

$$\nabla_{\mathbf{x}} \delta(c-f) = \nabla_{\mathbf{x}} f \partial_f \delta(x-f) = \nabla_{\mathbf{x}} f \partial_c \delta(x-f)$$

$$\nabla_{\mathbf{x}} \delta(\dot{c} - \nabla_{\mathbf{x}} f \cdot (\mathbf{p}-\mathbf{A})/m) = \nabla_{\mathbf{x}} (\nabla_{\mathbf{x}} f \cdot (\mathbf{p}-\mathbf{A})/m)$$

$$\partial_{\dot{c}} \delta(\dot{c} - \nabla_{\mathbf{x}} f \cdot (\mathbf{p}-\mathbf{A})/m)$$

$$\nabla_{\mathbf{p}} \delta(\dot{c} - \nabla_{\mathbf{x}} f \cdot (\mathbf{p}-\mathbf{A})/m) = \frac{1}{m} \nabla_{\mathbf{x}} f$$

$$\partial_{\dot{c}} \delta(\dot{c} - \nabla_{\mathbf{x}} f \cdot (\mathbf{p}-\mathbf{A})/m)$$

We have neglected possible surface terms while performing integration by parts. For example, if $x_i$ represents concentrations, one expects that $\rho(0,\mathbf{p}) \approx 0$ so mathematically one can extend the integration to $x \rightarrow -\infty$. Otherwise Eqn. 1 is not a good representation of the system dynamics in the first place. With $f_j = X_j = x_j$ Eqn. 21 reduces to Eqn. 9.

## E. Proof of GFDR for the linear system

To confirm the generalized fluctuation-dissipation relation, let's consider $t' \leq t$. We have,

$$< p_i x_i > = 0$$
$$< p_i^2 > = m$$
$$< x_i \xi_j(t) > = 0,$$
$$< \xi_i(t)\xi_j(t') > = (M_{ij} + M_{ji})\delta(t-t'),$$

$$< (\phi_{21}x_1 + \phi_{22}x_2)^2 > = \frac{\int d\mathbf{y} \left(\phi_{22} + \frac{\partial^2}{\partial x_2^2}\right) e^{-\frac{1}{2}\mathbf{y}^T \cdot \nabla\phi\nabla\phi \cdot \mathbf{y}}}{\int d\mathbf{y} e^{-\frac{1}{2}\mathbf{y}^T \cdot \nabla\phi\nabla\phi \cdot \mathbf{y}}} = \phi_{22}$$

Then,

$$\left< \left(\xi_1(t) - \frac{M_{12}}{M_{22}}\xi_2(t)\right)\left(\xi_1(t') - \frac{M_{12}}{M_{22}}\xi_2(t')\right) \right> = 2\left[M_{11} - \frac{M_{12}M_{21}}{M_{22}}\right]\delta(t-t')$$

$$\left< \left(\int_0^t d\tau e^{-\frac{\phi_{22}}{M_{22}}\tau}\left[\frac{\phi_{22}M_{12}}{M_{22}^2} - \frac{\phi_{12}}{M_{22}}\right]\xi_2(t-\tau)\right)\left(\int_0^{t'} d\tau' e^{-\frac{\phi_{22}}{M_{22}}\tau'}\left[\frac{\phi_{22}M_{12}}{M_{22}^2} - \frac{\phi_{12}}{M_{22}}\right]\xi_2(t'-\tau')\right) \right>$$
$$= -\left[\frac{M_{12}}{M_{22}} - \frac{\phi_{12}}{\phi_{22}}\right]^2 \phi_{22}\left[e^{-\frac{\phi_{22}}{M_{22}}(t'+t)} - e^{-\frac{\phi_{22}}{M_{22}}(t-t')}\right]$$

$$\left< \left(\xi_1(t') - \frac{M_{12}}{M_{22}}\xi_2(t')\right)\left(\int_0^t d\tau e^{-\frac{\phi_{22}}{M_{22}}\tau}\left[\frac{\phi_{22}M_{12}}{M_{22}^2} - \frac{\phi_{12}}{M_{22}}\right]\xi_2(t-\tau)\right) \right>$$
$$= -e^{-\frac{\phi_{22}}{M_{22}}(t-t')} \phi_{22}\left(\frac{M_{12}}{M_{22}} - \frac{\phi_{12}}{\phi_{22}}\right)\left(\frac{M_{12}}{M_{22}} - \frac{M_{21}}{M_{22}}\right)\left(1 - \frac{1}{2}\delta(t-t')\right)$$

$$\left\langle \left(\xi_1(t) - \frac{M_{12}}{M_{22}}\xi_2(t)\right)\left(\int_0^{t'} d\tau\, e^{-\frac{\phi_{22}}{M_{22}}\tau}\left[\frac{\phi_{22}M_{12}}{M_{22}^2} - \frac{\phi_{12}}{M_{22}}\right]\xi_2(t'-\tau)\right) \right\rangle$$

$$= -\phi_{22}\left(\frac{M_{12}}{M_{22}} - \frac{\phi_{12}}{\phi_{22}}\right)\left(\frac{M_{12}}{M_{22}} - \frac{M_{21}}{M_{22}}\right)\frac{1}{2}\delta(t-t')$$

$$e^{-\frac{\phi_{22}}{M_{22}}(t+t')}\left(\frac{M_{12}}{M_{22}} - \frac{\phi_{12}}{\phi_{22}}\right)^2 <(\phi_{21}x_1(0) + \phi_{22}x_2(0) + \phi_{23}x_3(0))^2> = e^{-\frac{\phi_{22}}{M_{22}}(t+t')}\left(\frac{M_{12}}{M_{22}} - \frac{\phi_{12}}{\phi_{22}}\right)^2\phi_{22}$$

By summing the above expressions, one has the generalized fluctuation-dissipation relation,

$$<F_1(t)F_1(t')> <p_1^2>^{-1} = K_{11}$$

### F. Projection with the quasi-linear system

Let's consider an $(N+1)$-dimensional quasi-linear system in the transformed representation,

$$0 = -\phi(x_0) - M_0\dot{x}_0 - \Phi_{0\mathbf{x}}\cdot\mathbf{x} - \mathbf{M}_{0\mathbf{x}}\cdot\dot{\mathbf{x}} + \xi_0(t)$$
$$0 = -\Phi\cdot\mathbf{x} - \mathbf{M}\cdot\dot{\mathbf{x}} - \Phi_{\mathbf{x}0}x_0 - \mathbf{M}_{\mathbf{x}0}\dot{x}_0 + \xi(t)$$

or in the original representation,

$$\frac{d}{dt}\begin{pmatrix} x_0 \\ \mathbf{x} \end{pmatrix} = \begin{pmatrix} M_0 & \mathbf{M}_{0\mathbf{x}} \\ \mathbf{M}_{\mathbf{x}0} & \mathbf{M} \end{pmatrix}^{-1}\cdot\begin{pmatrix} -\phi(x_0) - \Phi_{0\mathbf{x}}\cdot\mathbf{x} + \xi_0(t) \\ -\Phi\cdot\mathbf{x} - \Phi_{\mathbf{x}0}x_0 + \xi(t) \end{pmatrix}$$

We will project out the degrees of freedom $\mathbf{x} = (x_1,\cdots,x_N)$, and retain only the degree of freedom $x_0$. Notice that

$$\mathbf{x}(t) = e^{-\mathbf{M}^{-1}\cdot\Phi t}\mathbf{x}(0) + e^{-\mathbf{M}^{-1}\cdot\Phi t}\cdot\Phi^{-1}\cdot\Phi_{x0}x_0(0) + \int_0^t d\tau e^{-\mathbf{M}^{-1}\cdot\Phi \tau}\mathbf{M}^{-1}\cdot\xi(t-\tau) - \Phi^{-1}\cdot\Phi_{x0}x_0(t)$$

$$-\int_0^t d\tau e^{-\mathbf{M}^{-1}\cdot\Phi \tau}\left(\Phi^{-1}\cdot\Phi_{x0} + \mathbf{M}^{-1}\cdot\mathbf{M}_{x0}\right)\dot{x}_0(t-\tau)$$

$$\dot{\mathbf{x}}(t) = -\mathbf{M}^{-1}\cdot\Phi\cdot e^{-\mathbf{M}^{-1}\cdot\Phi t}\mathbf{x}(0) - e^{-\mathbf{M}^{-1}\cdot\Phi t}\cdot\mathbf{M}^{-1}\cdot\Phi_{x0}x_0(0) - \int_0^t d\tau e^{-\mathbf{M}^{-1}\cdot\Phi \tau}\cdot\mathbf{M}^{-1}\cdot\Phi\cdot\mathbf{M}^{-1}\cdot\xi(t-\tau) + \mathbf{M}^{-1}\cdot\xi(t)$$

$$+\int_0^t d\tau e^{-\mathbf{M}^{-1}\cdot\Phi \tau}\cdot\mathbf{M}^{-1}\cdot\left(\Phi_{x0} + \Phi\cdot\mathbf{M}^{-1}\cdot\mathbf{M}_{x0}\right)\dot{x}_0(t-\tau) - \mathbf{M}^{-1}\cdot\mathbf{M}_{x0}\dot{x}_0$$

Then,

$$0 = -\left\{\phi(x_0) - \Phi_{0x}\cdot\Phi^{-1}\cdot\Phi_{x0}x_0(t)\right\} - \int_0^t d\tau \dot{x}_0(t-\tau)\left\{2\left[M_0 - \mathbf{M}_{0x}\cdot\mathbf{M}^{-1}\cdot\mathbf{M}_{x0}\right]\delta(t)\right.$$

$$+\left[\mathbf{M}_{0x}\cdot\mathbf{M}^{-1}\cdot\Phi - \Phi_{0x}\right]e^{-\mathbf{M}^{-1}\cdot\Phi \tau}\cdot\mathbf{M}^{-1}\cdot\left(\Phi_{x0} + \Phi\cdot\mathbf{M}^{-1}\cdot\mathbf{M}_{x0}\right)\right\} + \left\{\xi_0(t) - \mathbf{M}_{0x}\cdot\mathbf{M}^{-1}\cdot\xi(t)\right.$$

$$+\left[\mathbf{M}_{0x}\cdot\mathbf{M}^{-1}\cdot\Phi - \Phi_{0x}\right]\cdot\int_0^t d\tau e^{-\mathbf{M}^{-1}\cdot\Phi \tau}\mathbf{M}^{-1}\cdot\xi(t-\tau) + \left[\mathbf{M}_{0x}\cdot\mathbf{M}^{-1}\cdot\Phi - \Phi_{0x}\right]e^{-\mathbf{M}^{-1}\cdot\Phi t}\mathbf{x}(0)$$

$$+\left[\mathbf{M}_{0x}\cdot\mathbf{M}^{-1}\cdot\Phi - \Phi_{0x}\right]e^{-\mathbf{M}^{-1}\cdot\Phi t}\cdot\Phi^{-1}\cdot\Phi_{x0}x_0(0)\right\}$$

The terms in the curled brackets are the derivative of potential of mean force, the memory kernel, and the random force terms, respectively.

## G. Numerical details

For the full model, the Langevin equations are propagated by

$$x_i(t_N) = x_i(t_{N-1}) + \Delta t G_i(\mathbf{x}) + \sqrt{2g\Delta t/\beta}\zeta_i(t),$$

where $\zeta_i(t)$ is generated from a Gaussian distribution with zero mean and unit variance, and we have set $\beta = 1$ throughout the work. We use $\Delta t = 0.005$ in all calculations. The potential of mean force is obtained from the steady-state distribution histogram. All the relaxation curves are averaged over 40000 trajectories.

To solve Eqn. 13 numerically, we first integrate both sides from $t_i = (i-1)\Delta t$ to $t_{i+1} = i\Delta t$,

$$0 = -\int_{t_{i-1}}^{t_i} dt' \frac{\partial}{\partial X_j} W(\mathbf{X}(t')) - 2\int_{t_{i-1}}^{t_i} dt' \Gamma_0 \dot{X} - \int_{t_{i-1}}^{t_i} dt' \int_0^{t'} ds\, 0.1pt \Gamma_1(s) \dot{X}(t'-s) + m\beta \int_{t_{i-1}}^{t_i} dt' F(t')$$

$$\approx -\frac{\partial}{\partial X_j} W(\mathbf{X}(t_{i-1}))\Delta t - \Delta t \Gamma_0 (X(t_i) - X(t_{i-1})) - \sum_{k=0}^{i} \Gamma_1((k+1/2)\Delta t)(X(t_{i-k}) - X(t_{i-k-1})) + m\beta \int_{t_{i-1}}^{t_i} dt' F(t')$$

We use the method of Berkowitz et al. to generate the random forces[37]. Then one realization of the random force is,

$$\int_{i\Delta t}^{(i+1)\Delta t} dt f(t) = \sum_{k=1}^{M} \left( \sqrt{\frac{J_K(2\pi k/P)}{P}} \right)$$
$$\left[ \frac{\zeta_{ak}}{\omega_k} \left( \sin(\omega_k(i+1)\Delta t) - \sin(\omega_k i\Delta t) \right) - \frac{\zeta_{bk}}{\omega_k} \left( \cos(\omega_k(i+1)\Delta t) - \cos(\omega_k i\Delta t) \right) \right]$$

where $\zeta_{ak}$ and $\zeta_{ak}$ are random numbers drawn from independent normal Gaussian distributions, $\omega_k = 2\pi k/P$, and $P = M\Delta t$ is the time interval that the random force does not repeat.

The spectral density is determined by the memory kernel through the Wiener-Khintchine theorem[18],

$$J_K(\omega) = 4\int_0^\infty dt\, \Gamma \cos(\omega t)$$

In all the simulations, we use $\Delta t = 0.1$, $M = 4000$, and all results are averaged over 40000 trajectories. We use the following fitting parameters for the memory kernel: $\gamma_0 = 28$; $\gamma_1$ is represented by four Gaussian functions in the form $\lambda_i \exp[-0.5((t-b_i)/a_i)^2]$, with lambda = (-5.2, -3, 1.56, 0.9), a=(1.5, 1.5, 1.5, 1.5), b = (8, 10, 16, 18).

The fitting procedure is as follows: first we use a piecewise linear function to obtain the analytical form of W from the stationary distribution; next we simulate the GLE with different sets of parameters of the memory kernel and $x_1(0) = 2<x_1>_{ss}$ to get the relaxation function, and choose the set that gives the minimum mean square error to the one calculated with the full model; then we run the GLE to generate the results shown in Fig 2 with the fixed set of parameters.

**Figure Caption**

Figure 1 Determination of the GLE parameters. (a) The network with end-product inhibition. (b) Potential of mean force calculated from the steady-state distribution of the full model. (c) The memory kernel used for fitting. (d) The fitted and simulated relaxation curve of $x_1$ with $x_1(0) = 2<x_1>_{ss}$. We did not fully optimize the fitting.

Figure 2 Comparison of the predicted and simulated relaxation functions with $x_1(0) / <x_1>_{ss}$ = 1.2 (a), 0.8 (b), 0.5 (c). The model parameters are the same as used in Fig.1.